\documentclass[aps,prb,reprint,amsmath,amssymb,superscriptaddress,footinbib,longbibliography]{revtex4-1}

\usepackage{siunitx}
\usepackage{textcomp}
\usepackage{graphicx}
\usepackage{gensymb}
\usepackage[utf8]{inputenc}
\usepackage{xcolor}

\begin{document}

\title{Ultra-high frequency vortex-based tweezers for microparticles manipulation with high spatial selectivity and nanoNewton forces}

\author{Roudy Al Sahely}
\affiliation{Univ. Lille, CNRS, Centrale  Lille, Univ. Polytechnique Hauts-de-France, UMR 8520, IEMN, F59000 Lille, 
France}%
\author{Jean-Claude Gerbedoen}
\affiliation{Univ. Lille, CNRS, Centrale  Lille, Univ. Polytechnique Hauts-de-France, UMR 8520, IEMN, F59000 Lille, 
France}%
\author{Nikolay Smagin}
\affiliation{Univ. Lille, CNRS, Centrale Lille, Univ. Polytechnique Hauts-de-France, UMR 8520 -
IEMN - Institut d’Électronique de Microélectronique et de Nanotechnologie, F-59000 Lille, France}%
\author{Ravinder Chutani}
\affiliation{Univ. Lille, CNRS, Centrale Lille, Univ. Polytechnique Hauts-de-France, UMR 8520 -
IEMN - Institut d’Électronique de Microélectronique et de Nanotechnologie, F-59000 Lille, France}%
\author{Olivier Bou Matar}
\affiliation{Univ. Lille, CNRS, Centrale Lille, Univ. Polytechnique Hauts-de-France, UMR 8520 -
IEMN - Institut d’Électronique de Microélectronique et de Nanotechnologie, F-59000 Lille, France}%
\author{Michael Baudoin}
\email{Corresponding author: \mbox{michael.baudoin@univ-lille.fr}}
\affiliation{Univ. Lille, CNRS, Centrale Lille, Univ. Polytechnique Hauts-de-France, UMR 8520 -
IEMN - Institut d’Électronique de Microélectronique et de Nanotechnologie, F-59000 Lille, France}%
\affiliation{Institut Universitaire de France, 1 rue Descartes, 75005 Paris}%

\begin{abstract}
Acoustical tweezers based on focused acoustical vortices open some tremendous perspectives for the in vitro and in vivo remote manipulation of millimetric down to micrometric objects, with combined selectivity and applied forces out of reach with any other contactless manipulation technique. Yet, the synthesis of ultra-high frequency acoustical vortices to manipulate precisely micrometric objects remains a major challenge. In this paper, the synthesis of a 250 MHz acoustical vortex is achieved with an active holographic source based on spiraling interdigitated transducers. It is shown that this ultra-high frequency vortex enables to trap and position individual particles in a standard microscopy environment with high spatial selectivity and nanoNewton forces. This work opens perspectives to explore acoustic force spectroscopy in some force ranges that were not accessible before.
\end{abstract}

\maketitle

\section{Introduction}
The precise manipulation of objects at small scales is paramount to exploring the properties of the micro- and nano-world. In the last decades, the development of optical and magnetic tweezers led to revolutions in microbiology \cite{arbbs_svoboda_1994}, such as the characterization of the dynamics and force of molecular motors \cite{nat_svoboda_1994,nat_finer_1996} or the exploration of the structural and mechanical DNA properties \cite{sci_smith_1992,sci_smith_1996,sci_strick_1996,bj_wand_1997}. Yet, these micro-manipulation techniques also suffer from major limitations: (i) The forces that these techniques can apply are typically limited to a few picoNewtons in the operating regime compatible with biomanipulations \cite{np_fazal_2011,sci_smith_1996,nsmb_dong_2010}, even if maximum forces in the nanoNewton range have been reported for specific anti-reflection coated particles and high beam power \cite{np_jannasch_2012}. (ii) Optical tweezers can induce photothermal and photochemical damages to living organisms \cite{Neuman1999,bj_liu_1995,bj_liu_1996,Blasquez2019} and cannot operate in vivo in optically opaque media. (iii) magnetic tweezers can only exert forces on magnetic beads, hence requiring pre-tagging of other particles and have low spatial selectivity owing to the low steepness of the trap. 

These limitations can be overcome by using single-beam selective acoustical tweezers \cite{arfm_baudoin_2020}, the acoustical analog of optical tweezers. Indeed, since the first theoretical and experimental demonstration of 3D selective trapping of a particle at the core of a focused acoustical vortex \cite{jap_baresch_2013,marzo2015,prl_baresch_2016}, it has been successively demonstrated, that acoustical tweezers (i) can safely and selectively manipulate individual cells \cite{nc_baudoin_2020} with forces of the order of 200pN, (ii) can operate in vivo \cite{pnas_ghanem_2020,pnas_lo_2021} or through complex opaque media such as the skull \cite{prap_jimenez_2019,r_yang_2021} and (iii)  can trap a variety of objects ranging from various particles \cite{marzo2015,prl_baresch_2016}, to cells \cite{nc_baudoin_2020} and bubbles \cite{pnas_baresch_2020}. 

Yet, these recent developments have long been hindered by the following paradox: On the one hand, spatial selectivity --- i.e. the ability to manipulate one particle independently of its neighbors -- and 3D trapping capability require (i) spatial localization and hence focusing of the acoustic energy to only affect the target particle and (ii) strong axial gradients to overcome axial scattering forces. But on the other hand, 3D selective trapping of micro-objects denser and stiffer than the surrounding medium cannot be achieved with a focused beam as in optics \cite{prap_gong_2022}. Indeed, this type of particles are generally expelled from the focal point of a focused beam, and only 2D trapping can be achieved near particular resonances of the trapped particle  \cite{prap_gong_2022}. This paradox has been solved by using specific beam structures called focused acoustical vortices \cite{jap_baresch_2013}. These beams are spinning around a phase singularity, wherein the pressure amplitude cancels, surrounded by a high intensity ring \cite{arfm_baudoin_2020}.

Since the first cylindrical acoustical vortex synthesis by Hefner and Marston in 1999 \cite{hefner1999}, many techniques have been developed to synthesize cylindrical and focused acoustical vortices: (i) active techniques based on arrays of bulk \cite{thomas2003,prl_volke_2008,courtney2014,marzo2015,prl_baresch_2016,r_yang_2021,ieee_hu_2021}  or surface acoustic wave transducers \cite{prap_riaud_2015,pre_riaud_2015,Riaud2016}, (ii) passive techniques based on helicoidal sources \cite{hefner1999}, photo-acoustics \cite{jasa_gspan_2004}, phase engineered surfaces \cite{ieee_ealo_2011,pp_muelas_2015,nat_melde_2016,mupb_terzi_2017}, metamaterials \cite{prl_jiang_2016}, space coiled paths \cite{prb_esfahlani_2017}, delay lines \cite{apl_marzo_2017} and diffraction gratings \cite{pp_jimenez_2015,pre_jimenez_2016,apl_jimenez_2018,apl_muelas_2018,ius_muelas_2021}. But all these techniques are strongly limited in the activation frequency and hence the dimension of the acoustical vortex. Yet, reducing the spatial dimensions of the acoustical vortex is mandatory to extend the capabilities of acoustical tweezers to smaller and smaller particles with increased selectivity and trapping forces, since these two last parameters strongly rely on the size of the vortex compared to the size of the particle \cite{nc_baudoin_2020}. Indeed, the vortex ring is repulsive for surrounding particles. Hence high selectivity can only be achieved when the vortex ring has a size close to the particle size. Additionally, strong trapping forces can only be obtained at high frequencies for small objects as they depends on the gradient of the acoustic field.

 In 2017 Riaud et al. \cite{prap_riaud_2017} proposed to use active holograms based on spiraling interdigitated transducers (IDTs) to synthesize cylindrical  and then focused acoustical vortices \cite{sa_baudoin_2019,nc_baudoin_2020}. They were able to synthesize 5 MHz and then 40 MHz acoustical vortices to trap microparticles and cells respectively. In theory, this technique could be extended to higher frequency, since it has been demonstrated that it is possible to synthesize surface acoustic waves with IDTs up to several hundred MHz \cite{am_shilton_2014} and even several GHz frequency \cite{ieee_brizoual_2008,rmp_friend_2011}. Nevertheless, this frequency increase is a challenging task since (i) the resolution required to achieve such high frequency approaches the limit of classical photolithographic techniques, (ii) these high frequency waves are extremely sensitive to the interfaces that they encounter and (iii) even the measurement of acoustical vortices at these time and spatial scales becomes challenging.
  
In this work, a 250 MHz acoustical vortex with a central ring of spatial extension \SI{9}{\micro\metre} in radius is synthesized with acoustical tweezers based on a spiraling IDT and characterized with an ultra-high frequency laser doppler vibrometer. With these tweezers, the selective manipulation of glass beads of \SI{4}{\micro\metre} in radius is demonstrated with forces in the nanoNewton range and conditions (temperature) compatible with biological manipulations. Physical understanding of the vortex synthesis is achieved with numerical simulations combining finite element (FEM) direct numerical simulation of the wave synthesis system and angular spectrum (AS) propagation of both the longitudinal and tangential waves in the solid \cite{jasa_boumatar_2022}. This mixed FEM/AS method enables the treatment of this high-frequency complex problem within a reasonable simulation time.

\begin{figure*}[htbp]
\centering
\includegraphics[width=\textwidth]{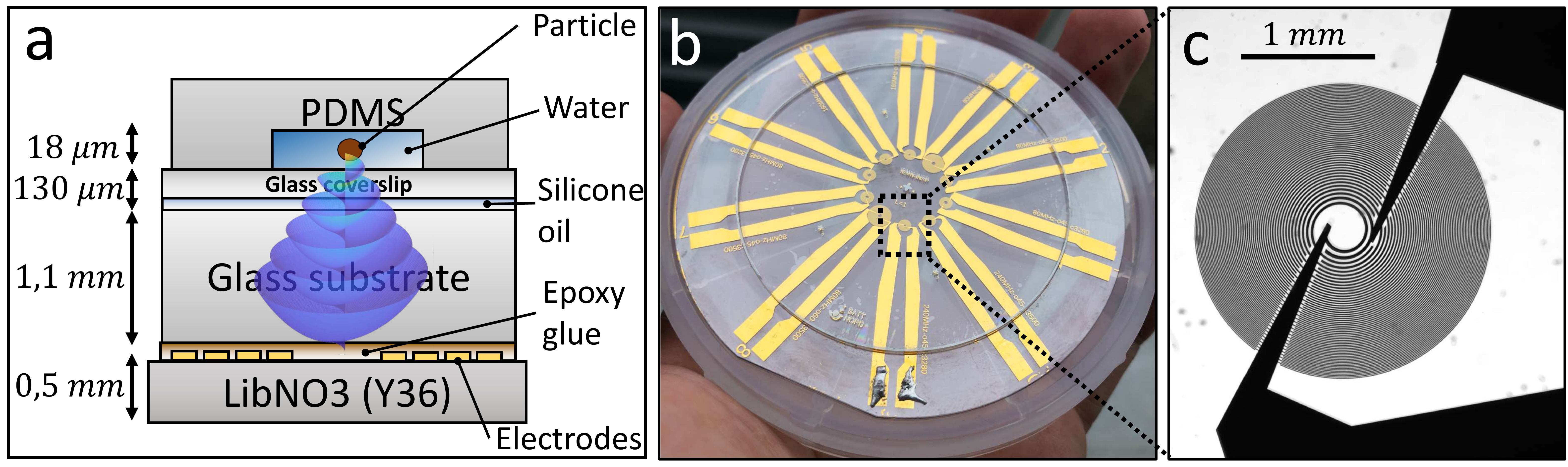}
\caption{Acoustical tweezers based on spiraling IDTs. a. Sectional sketch representing the different layers constituting an acoustical tweezers. b. Picture of a set of ten acoustical tweezers designed on top of a 3 inch Y36 Niobate Lithium wafer. c. Microscope view of the spiraling electrodes of the 250 MHz acoustical tweezers fabricated for this work.}
\label{figure1}
\end{figure*}

\section{Method}

\subsection{Tweezers fabrication}

The ultra-high frequency  tweezers are made of spiraling metallic electrodes sputtered at the top of a \SI{0.5}{\milli\metre} thick, 3 inches, Y-36 lithium niobate piezoelectric substrate (see sketch in Fig. \ref{figure1}). A \SI{1.1}{\milli\meter} borosilicate glass slide is glued on top of it to ensure the vortex focalization before the wave reaches the manipulation chamber. The particles to be manipulated are placed inside a microfluidic polydimethylsiloxane (PDMS) chamber sealed on top of a \SI{130}{\micro\metre} glass coverslip, acoustically coupled with the tweezers by a thin layer of silicone oil.  The tweezers  were fabricated  using the following process: 1. The shape of the spiraling electrodes are designed by using the equations provided in \cite{sa_baudoin_2019} and the code provided in \cite{nc_baudoin_2020} with the following parameters: driving frequency of 240 MHz, 30 turns, transverse wave speed $c_t = 3280 m s^{-1}$ and focal point corresponding to the top of the glass coverslip. 2. A high resolution chromium optical mask of these electrodes is prepared with an e-beam writer. 3. The spiraling metallic electrodes are deposited on top of the piezoelectric substrate using the following photolithographic technique: (i) LibNO$_3$ wafer is subjected to 3 minutes ultrasonication in acetone and propanol and dried with nitrogen gas. (ii) HMDS is spread on the wafer as an adhesion promoter using a spin coater, followed by a layer of negative AZnLoF2020 photoresist making a thickness of 2.9 $\mu$m. (iii) The resist is cured by placing the wafer on a hot plate at 110$^o$C for 90 seconds. (iv) The patterns of the tweezers are transferred into the resist by using the optical mask and MA6/BA6 SUSS Microtec UV optical aligner. (v) The wafer is placed on the hot plate at 110$^o$C for 120 seconds to finish the cross-linking process, and immersed in AZ326 developer for 30 seconds and then rinsed with deionized water. (vi) The wafer is coated with titanium (\SI{40}{\nano\metre}) and gold (\SI{400}{\nano\metre}) by evaporation. (vii) Lift off is achieved using SVC14 solution at 100$^o$C. (viii) The substrate is cleaned with propanol and dried using nitrogen gas.  4. A glass wafer of borosilicate D263 T (PGO Online) with a diameter of \SI{56.8}{\milli\metre}  and a thickness of 1.1mm is glued on the top of the piezoelectric substrate using optically transparent epoxy glue (EPOTEK 301-2). To ensure a good transmission of the wave, this step is essential. Note that prior to the glue step, a 15-nm-thick chromium layer is deposited by evaporation on the upper face of the glass wafer without etching. It will serve for the markers localizing the vortex center. The two substrates are cleaned with acetone and propanol followed by a O$_2$ plasma treatment to make the surfaces hydrophilic and improve the glue spreading. The epoxy glue is degassed using a vacuum box to prevent the formation of bubbles.  A drop of \SI{4.45}{\micro\liter} glue is added at the center of the piezoelectric substrate. The glass wafer is positioned on the top of the piezoelectric substrate and left on horizontal plate until the glue covers the whole surface between the lithium niobate and the glass substrates. After making sure that the whole surface is covered with glue, the latter is left to cure on the plate for 2 days at room temperature. 6. The last step is to etch the markers, which enable to identify the center of the vortex on the glass wafer in the microscope. To do so, a backside alignment process photo-lithography is performed. Thus, the whole structure is cleaned with acetone and propanol and dried with nitrogen gas to remove the dust.  The glass wafer is coated with AZ1505 resist (thickness about 0.5 $\mu$m) using a spin coater and placed on a hot plate at 110$^o$C for 90 seconds. After transferring the patterns on the resist using the optical mask and the MA6/BA6 SUSS aligner, the structure is placed in MIF726 developer and then rinsed with deionized water.  Cr etchant solution is used to remove the chromium layer. Finally, the tweezer is placed in acetone to remove resist traces followed by propanol and dried with nitrogen gas.

\subsection{Tweezers characterization}

Once the tweezers are fabricated, optical characterization of the vortex synthesized at the surface of the coverslip -- which supports the PDMS microfluidics chamber -- is carried out with a Polytec UHF-120 Laser Doppler Vibrometer (LDV). This heterodyne-type apparatus enables non-contact measurement of the out-of-plane vibrational component in the DC to 1.2 GHz frequency range. The acoustic tweezers covered with a \SI{130}{\micro\meter} glass coverslip acoustically coupled with the tweezers with a thin layer of silicone oil (25 cSt) were mounted on the LDV's motorized stage, allowing displacements in the 50 x 50 mm range with a \SI{0.3}{\micro\metre} resolution. The LDV performs spatial acoustic field scanning in point-by-point mode; a scan grid contains a total of 1599 (39 x 41) points with a 2.4 x 2.4 $\mu$m X-Y resolution. A Mitutoyo M Plan Apo 100x/0.70 infinity-corrected objective was used during the measurements. The 100x magnification ratio of this objective allows obtaining a 0.5 $\mu$m laser spot size and a 0.6 $\mu$m field depth, which ensures an accurate acquisition of the spatial field distribution. 
In order to obtain data in a wide frequency range of interest, the tweezers were excited with wave packets linearly swept from 220 to 270 MHz during 200 $\mu$s and a repetition period of 10 ms. These signals were amplified with an Amplifier Research 50W/1000A at a gain that provided 35 W of electrical power on a 50 Ohm load. Note that these high power are only used for the tweezers characterization to obtain a good signal-to-noise ratio and since in this case the tweezers are only activated 2\% of the time. Acoustic displacement data were acquired in the spectral domain, using 128000 fast Fourier transform (FFT) lines recorded at 1 Gs/s sampling rate during 320 $\mu$s; the corresponding frequency resolution is 3.125 kHz. Such an approach allows obtaining data on RMS of spatial field distribution (direct FFT data) and time-resolved images of vortex field (using the inverse FFT) at any frequency in the excitation range of 220-270 MHz. With this analysis we were able to determine that the optimal excitation frequency of the device corresponding to the maximum response was $250$ MHz, which was used as the excitation frequency in the experiments.

\subsection{Particle manipulation}

\begin{figure*}[htbp]
\centering
\includegraphics[width=\textwidth]{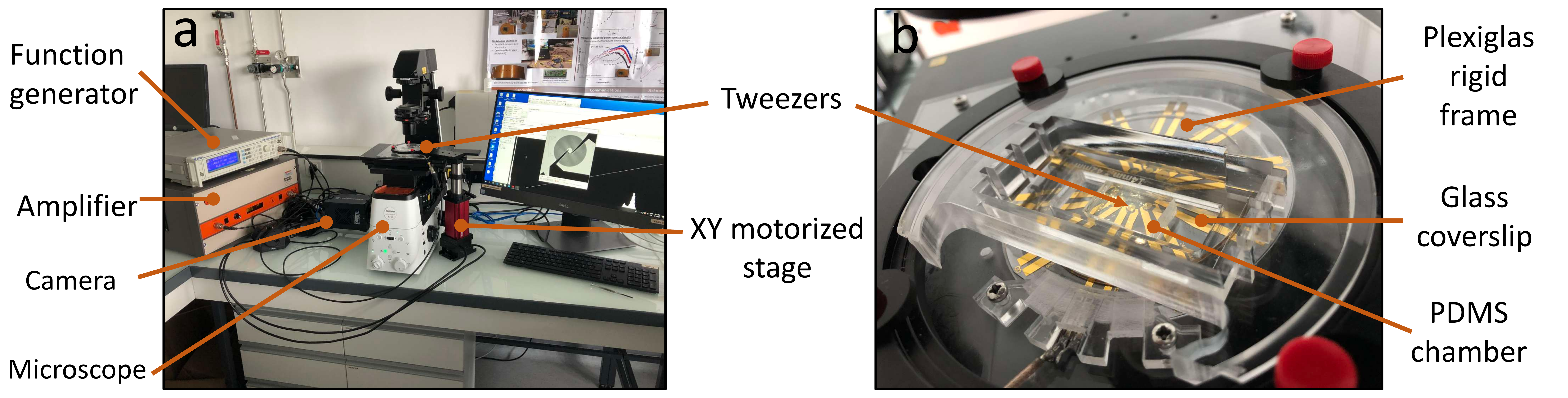}
\caption{Experimental setup. 
a. Picture showing the whole manipulation setup integrated in a Nikon Ti2E microscope. b. Zoom on the microfluidic chamber positioned on top of the acoustical tweezers with the rigid support used for the displacement of the chamber relative to the tweezers.}
\label{figure2}
\end{figure*}

For the manipulations tests, a drop of silica particles suspension (\SI{4}{\micro\meter} in radius) to be manipulated is pipetted onto the  \SI{130}{\micro\meter} thick glass coverslip and covered with a \SI{18}{\micro\meter} deep PDMS microfluidic chamber (see Fig. \ref{figure1}.a and \ref{figure2}.b). This coverslip is acoustically coupled to the tweezers with a thin layer of $25$ cSt silicone oil, which enables relative motion of the coverslip compared to the tweezers. This motion is achieved with a high precision (\SI{100}{\nano\meter}) XY Thorlabs PLS-XY motorized stage (see Fig. \ref{figure2}.a), which moves a rigid plexiglas frame glued to the glass coverslip. The whole setup is integrated into a Nikon Ti2E microscope, which enables visualization and recording of the particle displacement with a sCMOS Back Illuminated Prime-BSI photometrics camera at 40 fps. The tweezers are activated with a  IFR 2023A frequency generator driven at 250 MHz, whose signal is amplified with an AR50A250 150 W amplifier. The manipulation sequence is the following: (i) the particle to be manipulated is placed at the center of the markers localizing the vortex center by moving the microfluidic chamber with the motorized stage, (ii) the tweezers are activated with the electronics and (iii) the particle is moved with the motorized stage. Finally (iv)  the particle is released at the target position by switching off the power.

\subsection{Simulations}
\label{sec:simuls}

\begin{figure*}[htbp]
\centering
\includegraphics[width=\textwidth]{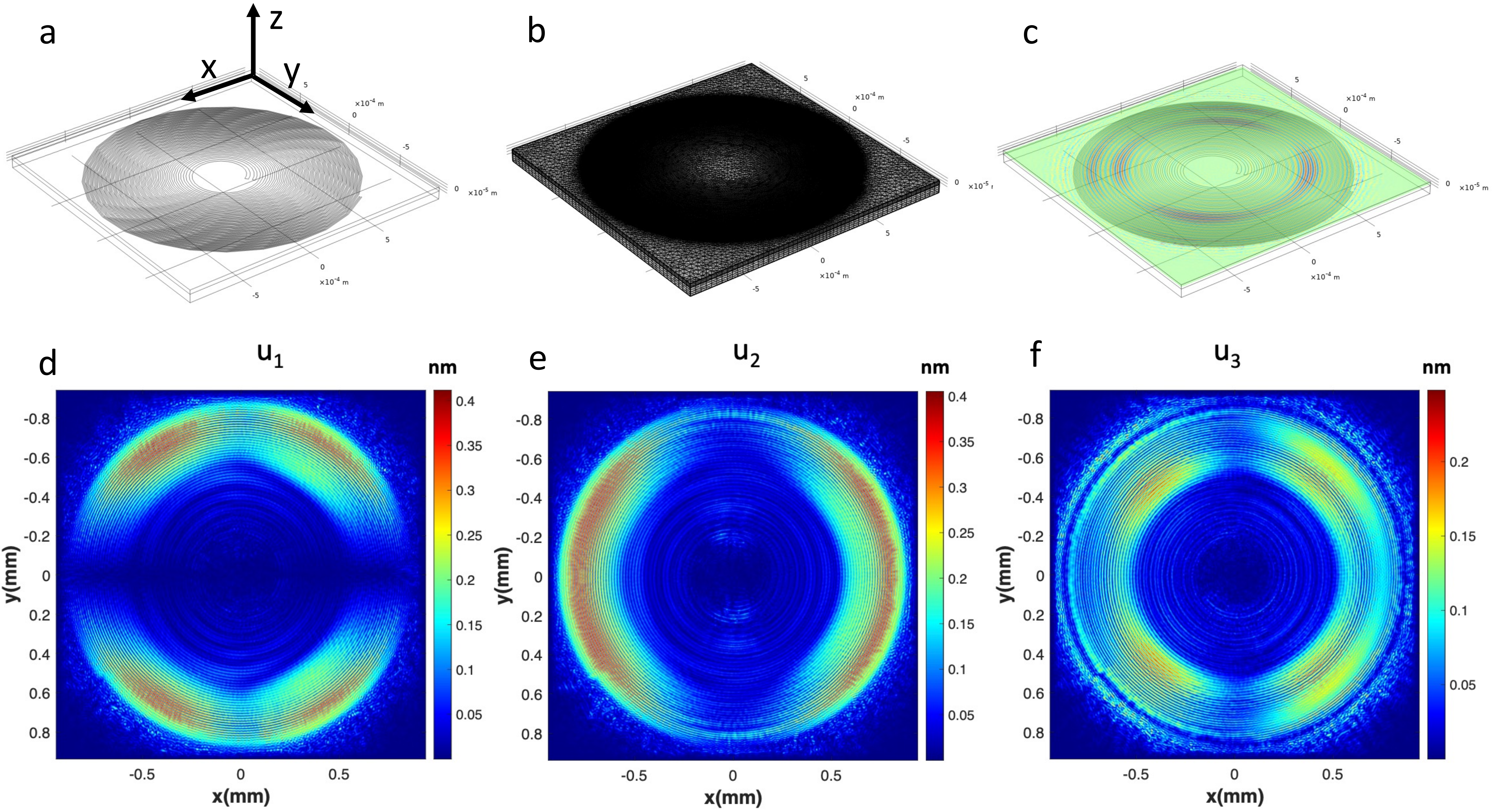}
\caption{Finite Element (FEM)  simulation of the wave synthesis. a. Geometry of the electrodes used for the simulations. b. Mesh used for the simulations. c. Real part of the normal displacement at the top of the piezoelectric substrate. d.e.f. Amplitude of the projections of the displacement field $u_1$, $u_2$ and $u_3$ over $x$, $y$ and $z$ directions respectively at the bottom of the glass substrate.}
\label{figure3}
\end{figure*}

To gain further insight of the active hologram wave synthesis physics, some simulations were performed with a mixed method \cite{jasa_boumatar_2022} combining: (i) a finite element (FEM) direct numerical simulation of the piezoelectric source, (ii) an angular spectrum (AS) technique propagating both longitudinal and tangential waves produced by the source in the glass slide and coverslip and then the longitudinal wave in the fluid contained in the microfluidic chamber, (iii) a proper use of transmission coefficients at each interface, and finally (iv) some perfectly matched layers to reduce the computation domain. This method enables to treat this 3D high frequency complex problem -- which would be hardly tractable with direct numerical simulation of the complete problem -- with a reasonable simulation time.

In short, the source consisting of the spiral-shaped metal electrodes (see Fig. \ref{figure1}.c) deposited at the surface of the LiNbO3 wafer is simulated using a commercial FEM software (COMSOL$^\copyright$  Multiphysics). The electrodes are considered to be infinitely thin and are represented in the numerical model as surface conditions for the electric potential. In order to limit the size of the finite element simulation, Convolutional Perfectly Matched Layers (C-PML) are used to absorb the waves leaving the computational domain in all directions. Hence, only the 4 upper wavelengths of the LibNO$_3$ and 2 lower wavelengths of the glued glass slide are simulated. In this way, the calculation domain is limited to 130 x 130 x 6 wavelengths. The C-PMLs occupy 3 wavelengths on each side of the computational domain, except on the top surface of the glass where they are only one wavelength of thickness. These finite element simulations enable the calculation of the three components of the displacement in the glass at one wavelength of the interface between LiNbO3 and the glass, when a potential of 10V is applied to one of the two electrodes, the other being grounded. The geometry of the electrodes used for the simulations, the mesh and the real part of the resulting vertical displacement $\mathbf{u_3}$ are shown in Fig. \ref{figure3} a, b and c respectively. The computaional domain is meshed using tetrahedra whose size does not exceed half a wavelength of the transverse wave in the glass. 

The displacement obtained by the finite element simulation is then decomposed into (i) a shear horizontal (SH) wave whose displacement is horizontal (parallel to the piezo/glass and glass/water interfaces), (ii)  a shear vertical (SV) wave and (iii) a longitudinal wave on the other hand. Since SH waves are transmitted in water only to a very thin viscous boundary layer of the order of $\delta_v = \sqrt{\mu}{\rho f} \sim$ \SI{64}{\nano\meter}, with $\mu$ the fluid viscosity, $\rho$ the fluid density and $f$ the frequency, they are not further considered in the fluid. The SH, SV and longitudinal waves are propagated to the interface between the glass layer and the water using the angular spectrum technique. The SV and longitudinal waves are then multiplied by the corresponding solid/fluid transmission coefficient to obtain the pressure in the fluid at the interface between the glass layer and the water. The pressure in the water is finally propagated in the thin fluid layer using also the angular spectrum technique. More details about the simulation method can be found in \cite{jasa_boumatar_2022}.

\section{Results and discussion}

\subsection{Vortex synthesis physics: numerical simulations and experiments}

\begin{figure*}[htbp]
\centering
\includegraphics[width=\textwidth]{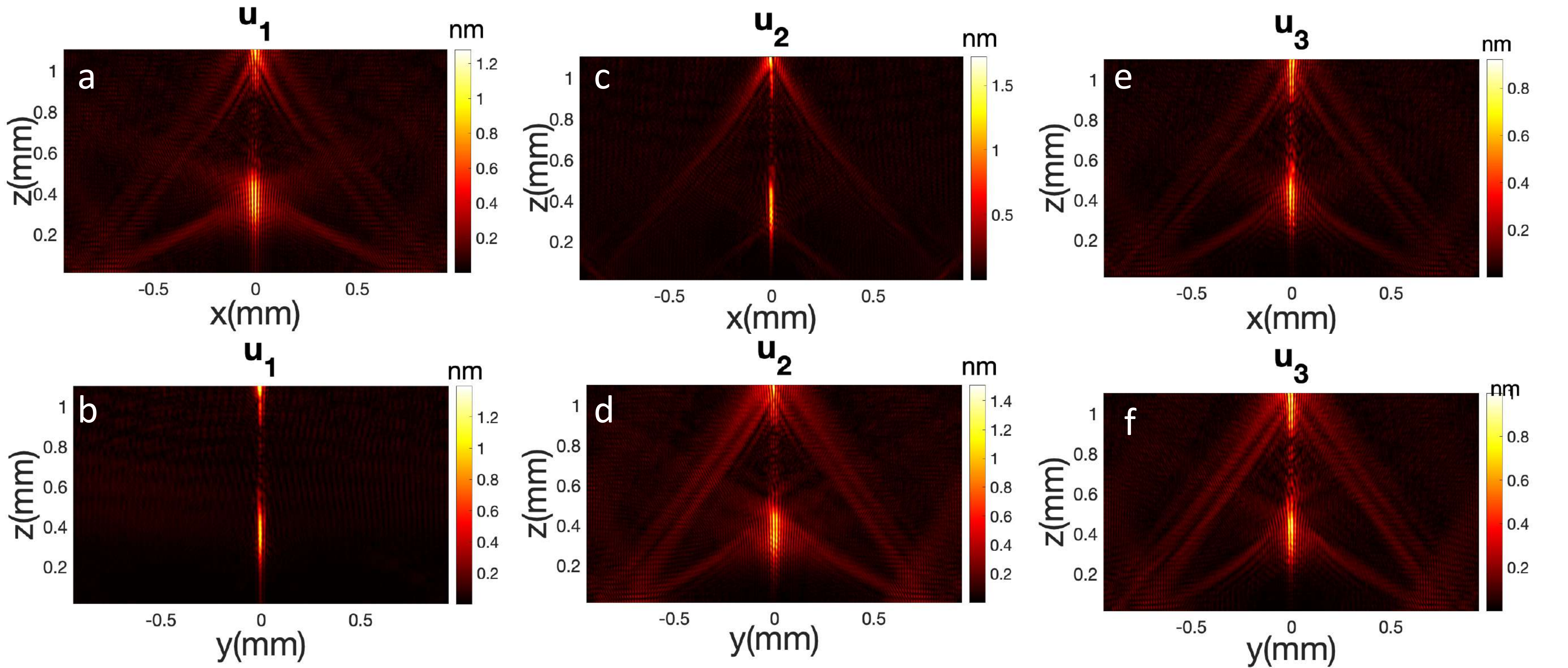}
\caption{FEM/AS simulation of the propagation of the three components of the displacement field $u_1$, $u_2$ and $u_3$ (in the $x$, $y$ and $z$ directions respectively) in the glass substrate and coverslip. The figure shows two focal points: one in the middle of the glass and one at the top of the coverslip.}
\label{figure4}
\end{figure*}

\begin{figure*}[htbp]
\centering
\includegraphics[width=0.7\textwidth]{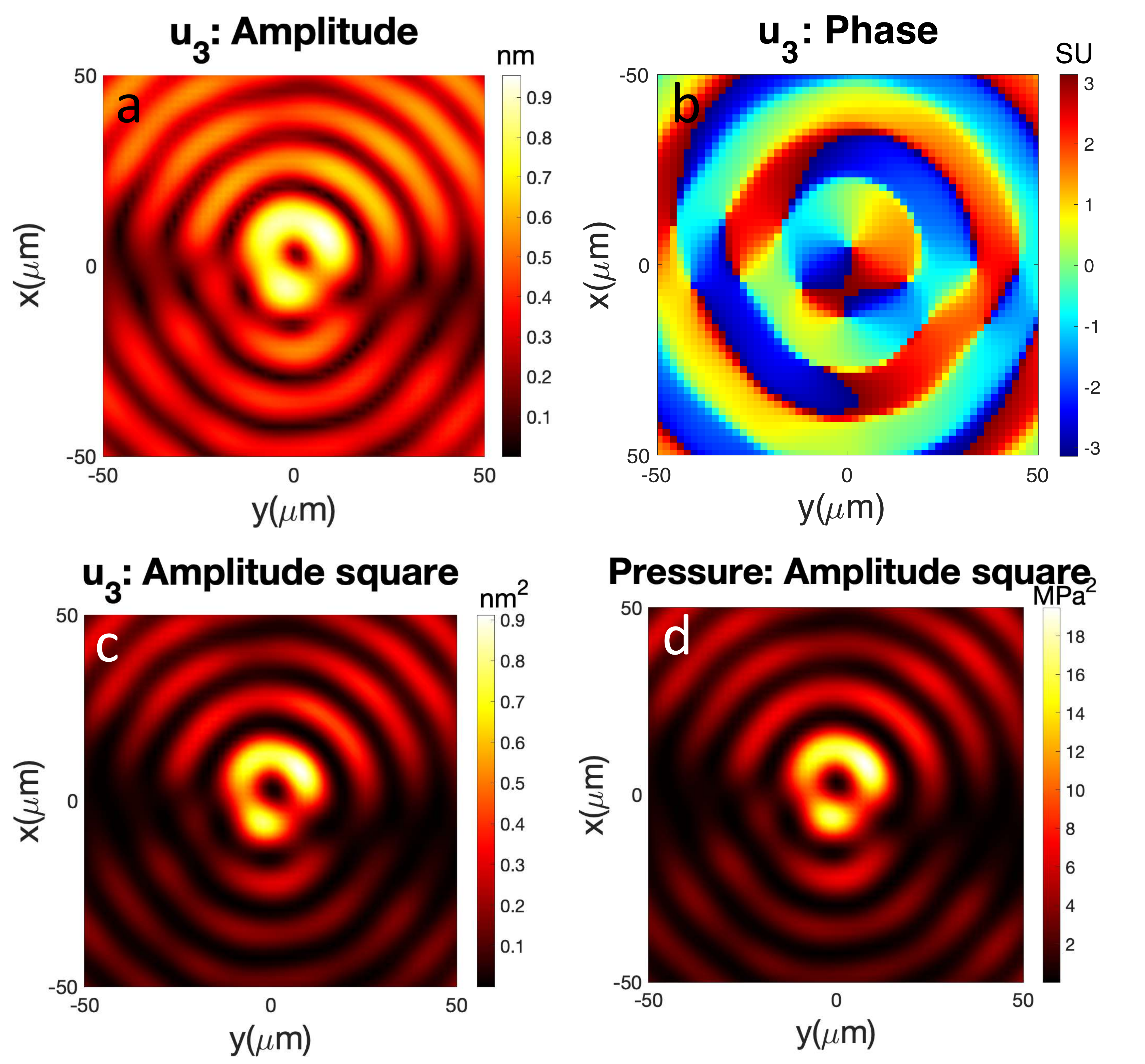}
\caption{FEM/AS simulation of the wavefield in the focal plane. a. Amplitude, b. phase and c. amplitude square of the normal displacement $u_3$ at the top of the coverslip. d. Pressure field in the fluid just above the coverslip.}
\label{figure5}
\end{figure*}

\begin{figure*}[htbp]
\centering
\includegraphics[width=0.75\textwidth]{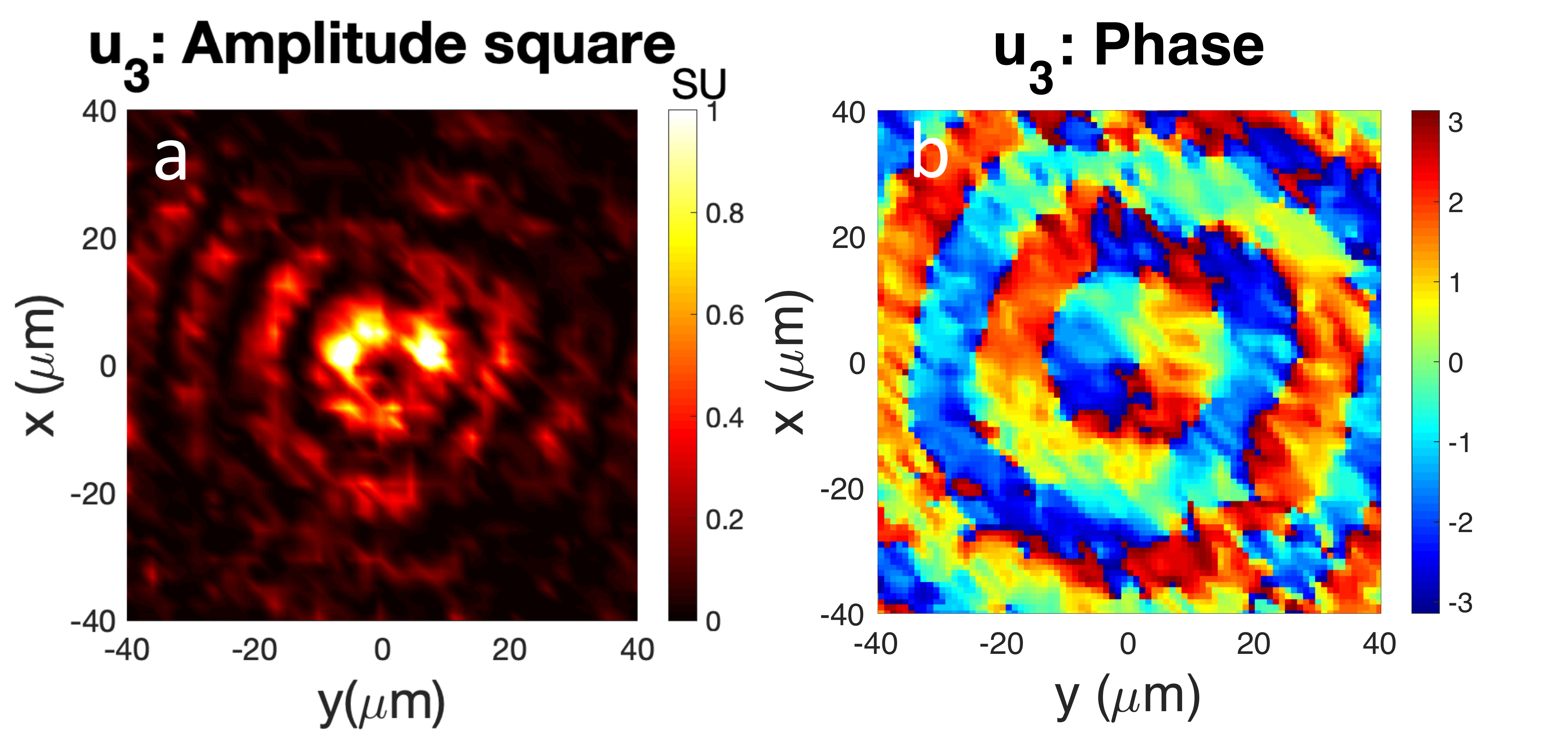}
\caption{Experimental measurement with a laser doppler vibrometer of the normal displacement at the top of the coverslip. a. Normalized amplitude square and b. Phase of the normal displacement $u_3$.}
\label{figure6}
\end{figure*}

The results of the simulations performed with the mixed FEM/AS code introduced in section \ref{sec:simuls} are presented in  Fig. \ref{figure3}, \ref{figure4} and \ref{figure5}. Fig. \ref{figure3} d, e, f represent the FEM simulated amplitude of the displacement $u_1$, $u_2$, $u_3$ (in the x,y,z directions respectively) in the glass resulting from the activation of the electrodes. This figure shows a certain anisotropy and inhomogeneity of the amplitude of the normal field in the glass $u_3$ resulting from the activation of the electrodes, while glass is an isotropic homogeneous medium. This anisotropy might be thought as the result from the anisotropy of the electromechanical coupling coefficients of the LibNO$_3$ in the different directions. But other simulations \cite{jasa_boumatar_2022} performed with a system wherein the electrodes radiate their signal directly in water or in a glass with a strongly different wave speed exhibit a much weaker anisotropy. Hence a possible explanation for this anisotropy might be the existence of specific interface modes appearing at the interface between the LibNO$_3$ and the glass due to the proximity of the waves speed in the Niobate lithium and glass, which, in some specific conditions (orientation, electrodes distance) could lead to some local amplification of the produced signal. Fig. \ref{figure4} shows the propagation of the signal (displacement $u_1$, $u_2$, $u_3$) produced at the bottom of the glass substrate up to the upper surface of the glass coverslip. The simulations exhibit two focal points resulting  from the activation of the electrodes : one at the top of the glass coverslip, which is used in the experiments to manipulate the particles, and one in the middle of the glass substrate. The existence of these two focal points is the result of the excitation of both transverse and longitudinal waves in the glass. Indeed, the shape of the electrodes was calculated to obtain the focalization of a transverse focused vortex at the top of the glass coverslip. But some longitudinal waves can also be produced. Indeed, (i) we demonstrated recently \cite{nc_baudoin_2020} that with the same electrodes, it is also possible to generate longitudinal waves with a focal point located at the top of the glass substrate, if it is actuated at a frequency $f_l = c_l / c_t \times f_t $, where $f_l$ is the actuation frequency for longitudinal vortex, $f_t$ the actuation frequency for the transverse vortex, $c_l$ is the longitudinal wave speed and $c_t $ the transverse wave speed and $c_l / c_t \approx 1.6$. But here the activation frequency is $f_t$. (ii) In another recent paper \cite{prap_gong_2021}, we demonstrated that the position of the focal point of a focused vortex generated by an active spiraling transducer, can be moved axially by tuning the excitation frequency. In particular, if the frequency is reduced, the focal point is moved downward closer to the source. This is exactly what is observed in Fig. \ref{figure4}. If the electrodes had been activated at $f_l = c_l / c_t f_t$, the longitudinal wave would have focused at the top of the glass coverslip. But since a much lower activation frequency $f_t$ is used, the focal point of the longitudinal wave is moved downward inside the glass. However, we can note that this longitudinal wave plays a minor role on the manipulation device. Indeed, after the focal point, the longitudinal wave diverges and will be scattered inside the glass, resulting in incoherent waves. The magnitude of this incoherent wave will remain weak compared to the intensity of the coherent transverse wave which impacts the glass's upper surface and is used for particle manipulation. It nevertheless can result in some noise, as observed in the experimental measurements (Fig. \ref{figure6} a and b). 
Fig. \ref{figure5} shows the normal displacement (amplitude, phase, amplitude square) expected at the surface of the glass from the FEM/AS code and inside the water contained in the microfluidic chamber. As expected, the normal displacement and pressure fields are very similar since only the normal displacement is transmitted to the fluid. The simulated vortex shows a certain degree of anisotropy, which simply results from the anisotropy observed just above the electrodes. Finally, Fig. \ref{figure6} shows the signal measured with the high frequency laser doppler vibrometer.  This figure confirms the synthesis of the high frequency vortex, with an amplitude and phase close to the simulations predictions and a trap size (radius of the first ring) of the order of 9$\, \mu m$.

\subsection{Particle displacement: Selectivity and nanoNewton force.}

\begin{figure}[htbp]
\centering
\includegraphics[width=0.5\textwidth]{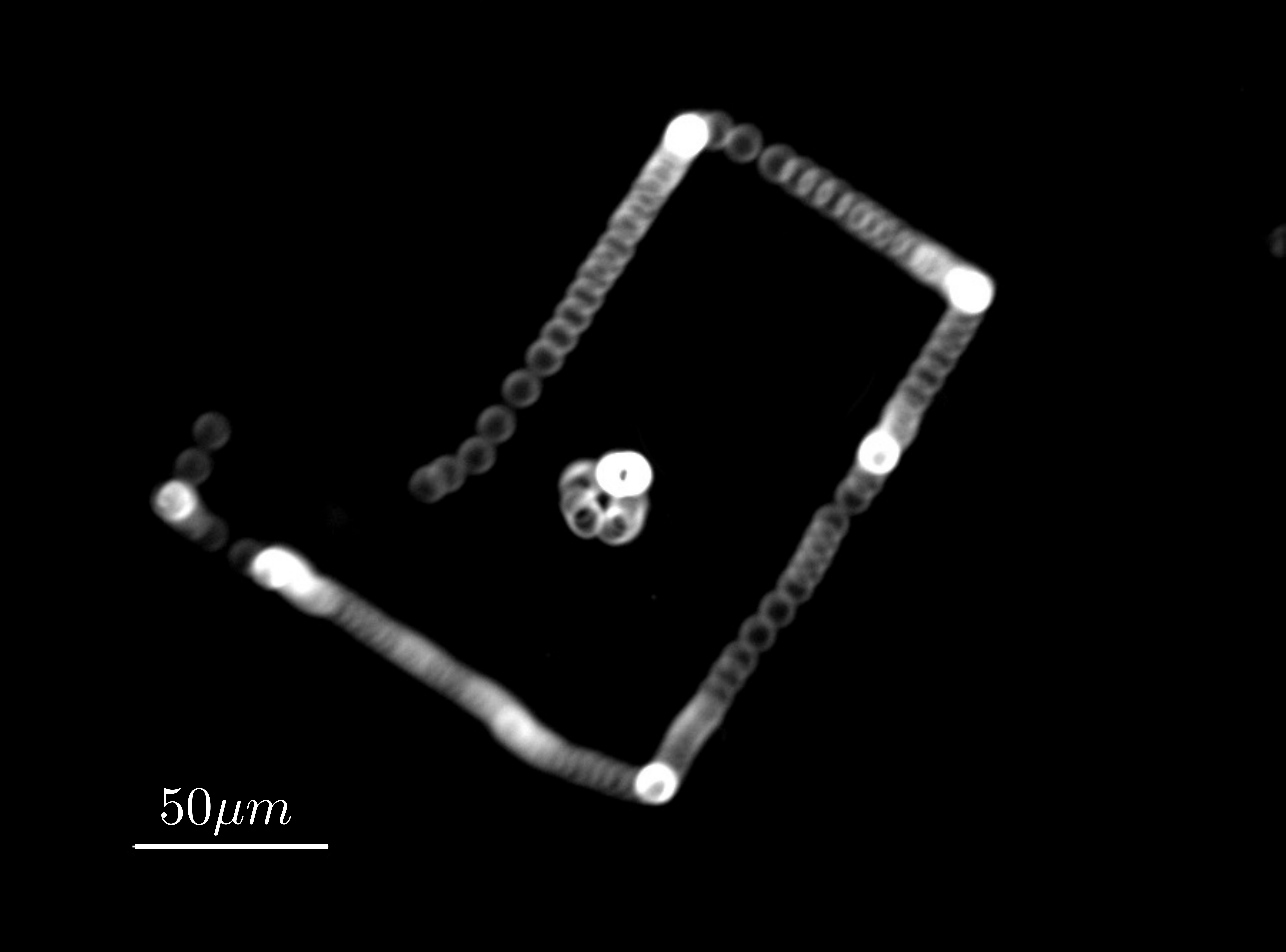}
\caption{Figure illustrating the precision of the displacement of a trapped \SI{4}{\micro\metre} silica particle around a reference particle to form a rectangle. Note that in the experiments, the particle which moves is the particle at the center of the rectangle and the reference particle is the one that forms the rectangle. Indeed, in the experiments the tweezers and hence the center of the trap remains fixed, and the coverslip is moved comparatively to the tweezers. In the treatment of the data, obtained by standard deviation of the superposition of the stack of images shot by the camera, only the trapped particle and reference particles, are kept, the other particles are erased. This figure corresponds to movie M2.}
\label{figure7}
\end{figure}

The ability of these high frequency tweezers to manipulate some $4 \, \mu m$ silica beds was then investigated. Movie M1 shows the displacement of a target particle in between other particles. Movie M2 illustrates the precision and selectivity of the tweezers: (i) a particle is selected and moved precisely around  another reference particle to form a rectangle (see Fig. \ref{figure7}) and then the reference particle is moved to show that particles can be selectively picked up and moved independently. In the experiments, we can notice a slight adherence of some particles, especially when the particles are introduced in the channel several minutes before the manipulation. We can note in Movie M2 and Fig. \ref{figure7} that when the particle executes the fourth side of the rectangle, it sometimes escapes from the trap and moves on the first dark ring before coming back to the center of the trap. This is only observed when the particle is moved in one specific direction. This is consistent with the anisotropy of the amplitude of the first ring observed both in the simulations (Fig. \ref{figure5} c and d) and experiments  (Fig. \ref{figure6} a), which create some weakness in the first ring wherein the amplitude of the wave, and hence the force of the trap, is weaker. Finally, we conducted some experiments to measure the trapping force of these ultra-high frequency tweezers. A particle is trapped and accelerated until the particle escapes from the trap. A maximum translation speed of $11$ $mm s^{-1}$ inside a \SI{18}{\micro\metre} height channel was measured when the particle reached the edge of the microchamber. The calculation of the force with Faxen's formula gives an estimation of the force of 1.4 nN. Note that the particle is lost because it reaches the edge of the chamber and not because the stokes drag made it escape from the trap. Hence, 1.4 nN constitutes a lower bound for the value of the trapping force. Note also that all the experiments were performed with driving signals leading to a temperature increase of less than 2.5\textdegree C during the manipulation (see SI), some conditions that remain compatible with biological manipulations.

\section{Conclusion}

In this paper, we demonstrate that the method of active spiraling holograms enables to design highly selective ultra-high frequency selective tweezers. The field generated at the bottom of a microfluidic chamber is characterized with a laser Doppler vibrometer and successfully compared to numerical predictions obtained with a mixed FEM/AS code. This code enables the computation of this ultra high frequency problem with a reasonable computation time. With these tweezers, it is demonstrated that $4 \mu m$ particles can be picked up and moved selectively with forces in the nanoNewton range. In future work, it would be interesting to investigate (i) ways to reduce the anisotropy of the vortex and obtain a more homogeneous trap and (ii) further optimization of the piezoelectric material (e.g. the LibNO$_3$ cut). Such an optimization would enable to create signals of higher intensity with the same input power, e.g. by increasing the synthesis of the transverse wave and reducing the synthesis of the longitudinal wave.

\section*{Acknowledgements}
This work was supported by ISITE-ULNE (Prematuration and ERC Generator programs), Institut Universitaire de France and Renatech Network. We acknowledge the platform WaveSurf from Université Polytechnique Hauts de France for performing the measurements with the laser Doppler vibrometer.

\end{document}